\documentclass[aps, twocolumn]{revtex4-1}

\usepackage{graphicx}
\usepackage{amsmath}
\usepackage{dsfont}
\usepackage{amssymb}
\usepackage{physics}
\usepackage{hyperref}
\usepackage{ulem}
\hypersetup{colorlinks=true,
	    final=true,
	    linkcolor=blue,
	    citecolor=blue,
	    filecolor=blue,
	    urlcolor=blue,}

\usepackage{amssymb}

\newcommand{\pr}{\partial}

\newcommand{\ep}{\epsilon}

\newcommand{\p}{\prime}

\newcommand{\om}{\omega}

\newcommand{\beq}{\begin{equation}}
\newcommand{\eeq}{\end{equation}}

\newcommand{\mathq}{\mathbf{q}}
\newcommand{\mathk}{\mathbf{k}}
\newcommand{\mathkp}{\mathbf{k^\p}}

\newcommand{\vtb}{\bar{v}_t}

\begin{document}

\title{Kaganov-Lifshitz-Tanatarov theory for tilted Dirac-cone materials: anisotropic heating form uniform light}

\author{Navinder Singh$^{1,*}$, Bharathiganesh Devanarayanan$^1$, Sruthi Sudhakaran$^2$, Jalaja Pandya$^1$, and Saptarshi Mandal$^3$}

\affiliation{$^1$Theoretical Physics Division, Physical Research Laboratory (PRL), Ahmedabad, India. PIN: 380009. $^2$Department of molecules and materials, University of Twente, Enschede, 7522NB, The Netherlands. 
$^3$Institute of Physics, Bhubaneswar 751005, Odisha, India}

\begin{abstract}
We point out that in the tilted Dirac cone materials the non-equilibrium (hot) electron relaxation with phonons is anisotropic in the Brillouin zone. It means that there is a preferential heating of the lattice degrees of freedom in the specific directions of the Brillouin zone, in particular, in the direction opposite to the tilt velocity in the model considered by us.  This observation will have novel consequences: (1) With pump-probe spectroscopy applied to a given tilted Dirac cone material an anisotropic relaxation would lead to a transient anisotropic heating which can further lead to a transient Seebeck effect as transient thermal gradients would exist in the specific directions of the BZ, and (2) this direction of anisotropic heating can be controlled by controlling the direction of the tilt velocity which can be externally tuned by the application of an external pressure. We foresee novel applications of this effect in ultrafast sensor applications involving transient heating effects. This is equivalent to inducing a transient Seebeck effect by just shinning light on a tilted Dirac cone material!
\end{abstract}

\email{Corresponding author: Navinder Singh (navinder.phy@gmail.com; navinder@prl.res.in)}

\maketitle

 \section{Introduction} 
 
Non-equilibrium electron relaxation in tilted Dirac cone materials has not been fully explored.  In this work we initiate the topic of non-equilibrium electron relaxation in tilted Dirac cone materials through an extension of the Kaganov-Lifshitz-Tanatarov (KLT) theory. To introduce the problem let us imagine a sheet of a tilted Dirac-cone material (such as PdTe$_2$ or 8-Pmmn Borophene). Imagine that its surface is illuminated with a femto-second pump pulse. Assume that the dimensions of the area illuminated is around $1~\mu m$. Within this area, the pump pulse would generate a non-equilibrium state of hot electrons that would relax via phonon generation. Below we show that this phonon generation is anisotropic in a tilted Dirac-cone material and in the area illuminated, a transient  thermal gradient will set up which will have non-trivial consequences. Before we explain this effect, we briefly review the  two-temperature model (TTM) of hot electron relaxation in metals which provides the foundation on which this new work is built. 
 
The two-temperature model (TTM) of hot electron relaxation in metals was first developed while studying radiation damage caused by very energetic ions\cite{klt}.  When such ions penetrate a metal, they lose energy mainly to the electrons via scattering. This quickly heats up the electron subsystem. Fast electron-electron scattering then creates a hot Fermi-Dirac distribution at an elevated electron temperature (elevated with respect to the lattice temperature). Later, this energy is transferred more slowly to the lattice through electron-phonon scattering. Thus it was assumed that the electron-electron relaxation is much faster than electron-lattice relaxation. During electron-phonon scattering, both the subsystems (electrons and phonons) assumed to remain in local thermal equilibrium. This is the foundation on which TTM was constructed. For details, refer to reviews\cite{nav1,nav2}. 
 
Very briefly, the Kaganov-Lifshitz-Tanatarov (KLT) theory of the Two-Temperature Model (TTM) can be expressed in the following way:

KLT assumed that after heating, electrons quickly regain a "hot" Fermi-Dirac distribution: 
\beq
f_k =\frac{1}{e^{\beta_e(\ep_k -\mu)} +1}, ~~~\beta_e =\frac{1}{k_B T_e},
\eeq
where $T_e$ is the temperature of the electron sub-system (greater than the lattice temperature $T$ during the process of relaxation). Free electron model was used: $\ep_k = \frac{\hbar^2k^2}{2 m}$. For phonons, Bose distribution was assumed:
\beq
n_q =\frac{1}{e^{\beta\hbar\om_q} - 1}, ~~~\beta =\frac{1}{k_B T},
\eeq
where $q$ is the magnitude of the wave-vector of an acoustic phonon mode. Debye model was used for phonon sub-system ($\om_q = c_s q$ where $c_s$ is the sound speed for acoustic phonons). During the process of relaxation heat transfers from the electron subsystem to phonon subsystem and then by the process of diffusion it goes out to the substrate or to the ambient medium. 

The amount of the average energy transferred by electrons to phonons per unit time and per unit volume is given by: 

\beq
\bar{U} = \int d^3r \frac{d^3q}{(2\pi)^3} \dot{N}_q \hbar\om_q.
\label{3}
\eeq
Where $\dot{N}_q$ is the phonon generation rate, and $\dot{N}_q \hbar \om_q$ is the amount of energy transferred (per sec per unit volume) to phonon modes with wave vector lying in the range $q$ to $q+dq$.  $\dot{N}_q$  is computed using the Bloch-Boltzmann equation:

\begin{eqnarray}
\dot{N}_q&=&2 \int \frac{d^3k^\p}{(2\pi)^3} W_{k,k'} f_{k'} (1-f_k) [(n_q+1)\nonumber\\
&\times&\delta(\ep_{k'} -\ep_k -\hbar\om_q)- n_q \delta(\ep_{k'} -\ep_k +\hbar\om_q)].
\end{eqnarray}

\beq
W_{k,k'} = \frac{\pi D^2}{\rho V c_s^2}\om_q,~~~~\mathq = \mathk - \mathkp.
\eeq
 
Here $D$ is the deformation potential constant, $\rho$ is the density of metal, $V$ is unit cell volume, and $c_s$ is the sound speed\cite{klt,nav1,nav2}. By using the expressions for Fermi and Bose functions and using the assumption $\mu>> \hbar \om_q$, the above equation takes the form: 

\beq
\dot{N}_q=\frac{m^2U^2 \hbar\om_q}{2\pi \hbar^4 \rho V c_s}\frac{e^{\beta \hbar \om_q} -  e^{\beta_e \hbar \om_q}}{ (e^{\beta \hbar \om_q} - 1) (e^{\beta_e \hbar \om_q} - 1)}
\eeq

Technical details are given in\cite{nav1}.  The average energy transferred in the low temperature limit $T, ~T_e<< T_D$  is given as:
\beq
\bar{U} \sim T_e^5 -T^5
\eeq

In the high temperature limit $T,~T_e>>T_D$, it turns out to be:
\beq
\bar{U} \sim T_e-T.
\eeq

This behaviour of energy relaxation within the TTM mimics the temperature dependence of the DC resistivity due to phonon scattering\cite{nav3,ziman}. 

In the next section we extend the KLT theory to the case of graphene. As far as we know, the direct extension of the KLT theory following the original method\cite{klt} has not been done in the literature. However, alternative methods based on the density matrix, time-dependent Boltzmann equation, and numerical solutions of the Bloch-Boltzmann equation applied to graphene exist, and our conclusions in appropriate limits match with those\cite{kristen,sarma,buts,mohd,caru}. In addition, the analytical solution would provide insight for the application of KLT theory for the tilted Dirac cone materials which is the main objective of this work (section III). In section (IV) we summarize our results.

\section{Kaganov-Lifshitz-Tanatarov theory for graphene}
 
To develop KLT theory for graphene we need to take into account the following basic facts:

  \begin{figure}[h!]
    \centering
    \includegraphics[width=0.6\columnwidth]{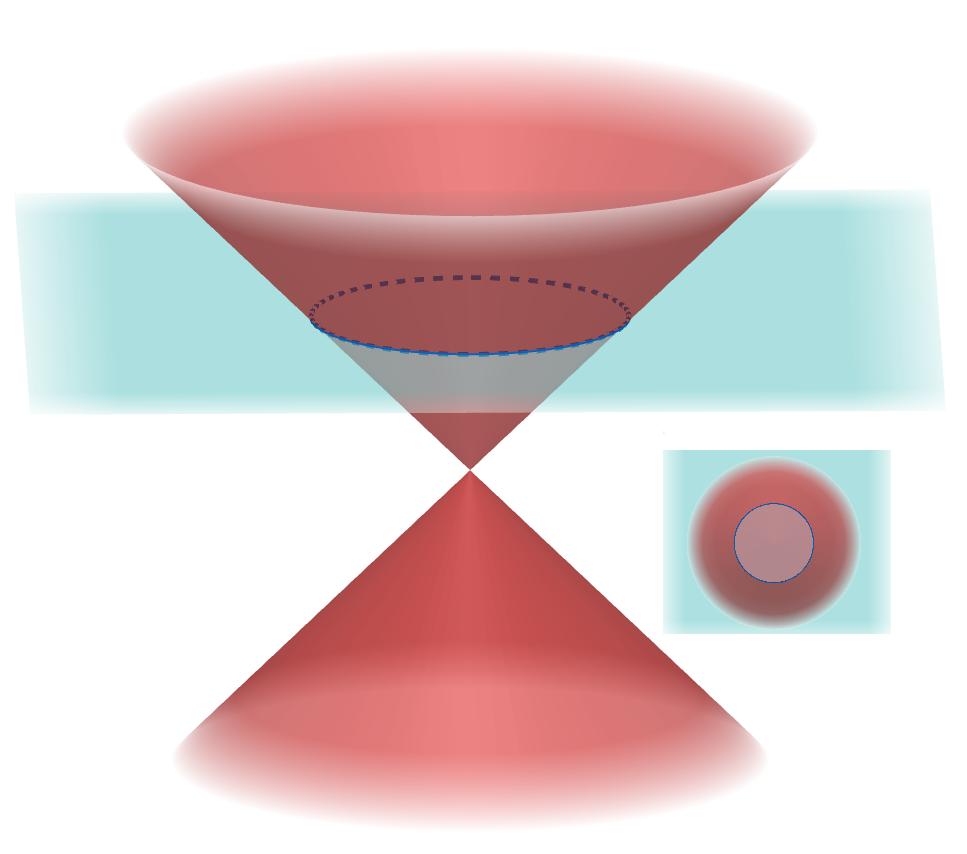}
    \caption{Dirac Cone (no tilt).}
  \end{figure}

\begin{enumerate}

\item For relevant energy scales, electronic dispersion in the Dirac cone is linear in wave vector:  $\ep_k = \hbar v_F |\mathk| = \hbar v_F k$. We assume that Fermi energy is tuned and lies in the conduction band.

\item We assume that the scattering rate due to electron -(acoustic)phonon interaction is given by\cite{sarma,enr}:
\beq
W_q = \frac{\pi D^2 q}{\rho_m c_s } \left( 1- \left(\frac{q}{2k}\right)^2\right).
\eeq
Here $D$ is the deformation potential constant, $c_s$ is the sound speed, and $\rho_m$ is the mass density $(kg/m^2)$ of graphene. 
\end{enumerate}

In the case of two-dimensions (for graphene sheet) the equation (4) reads:

\begin{eqnarray}
\dot{N}_q&=&2 \int \frac{d^2k^\p}{(2\pi)^2} W_{k,k'} f_{k'} (1-f_k) [(n_q+1)\nonumber\\
&\times&\delta(\ep_{k'} -\ep_k -\hbar\om_q)- n_q \delta(\ep_{k'} -\ep_k +\hbar\om_q)].
\end{eqnarray}

We next proceed to the computation of $\bar{U}$ (the rate of energy relaxation to phonons). Introduce momentum conservation (all vector are in 2D) and rearrange the integrand using properties of the delta function. The rate of phonon generation per unit area and per unit time is then given by

\begin{widetext}
\beq
\dot{N}_q=2 \int d^2k \int \frac{d^2k^\p}{(2\pi)^2} W_{k,k'} [f_{k'} (1-f_k) (n_q+1) - f_{k} (1-f_{k'}) n_q]\delta(\ep_{k'} -\ep_k -\hbar\om_q) \delta(\mathk' - (\mathk+\mathq)).
\eeq
$k'$ integral can be immediately done using the properties of the delta function over momenta. Fermi and Bose factor can be written in the following way:

\beq
f_{k'} (1-f_k) (n_q +1)-  f_{k} (1-f_{k'}) n_q = \left( \frac{1}{e^{\beta_e (\ep_k - \mu)}+1} - \frac{1}{e^{\beta_e (\ep_k+\hbar\om_q - \mu)}+1} \right)[n_q(\beta_e) - n_q(\beta)].
\eeq

Thus, the expression for $\dot{N}_q$ takes the form
 
 \beq
 \dot{N}_q=\frac{2}{(2\pi)^2}\frac{\pi D^2 q}{\rho_mc_s} \left( 1- \left(\frac{q}{2k}\right)^2\right) [n_q(\beta_e) - n_q(\beta)] \int d^2k  (f(\ep_k) - f(\ep_k+\hbar\om_q))  \delta(\ep_{k+q} -\ep_k -\hbar\om_q).
\eeq

\end{widetext}
 
Where we have used the expression for the scattering rate (equation(9)). We assume that the Fermi energy is the largest energy scale in the problem and the following case holds true: 

\beq
\hbar\om_q,~k_BT_e,~k_BT<<\mu.
\eeq 

With this, we can use the following Taylor expansion for the Fermi functions 

\beq
 f(\ep_k) - f(\ep_k+\hbar\om_q) \simeq -\hbar\om_q \frac{\pr f}{\pr\ep_k},
 \eeq
 and
 \beq
 -\frac{\pr f}{\pr\ep_k} \simeq \delta(\ep_k-\mu).
 \eeq
 
Collecting this information, the integral over $k$ in equation (13) can be simplified to 
 
\beq
\frac{1}{(\hbar v_F)^2}\frac{k_F}{q} \frac{1}{\sqrt{1-(q/2k_F)^2}}.
\eeq  
  
The calculation details are given in the appendix.  Equation (13) now reads

\beq
 \dot{N}_q=\frac{D^2 k_F}{2\pi \hbar\rho_m c_S^2}\left(\frac{c_S}{v_F}\right)^2 q \sqrt{1- \left(\frac{q}{2k_F}\right)^2} [n_q(\beta_e) - n_q(\beta)].
 \eeq

Next, the energy transferred per unit area per unit time is give as

\beq 
\bar{U} = \int \frac{d^2 q}{(2\pi)^2} \dot{N}_q \hbar\om_q
\eeq
Where the expression for $ \dot{N}_q$  is given by equation (18). To compute $\bar{U}$, it turns out that we need to deal with the following kind of integral over q:

\beq
I(\beta_i) = \int_0^{q_{BG}} dq q^3 \sqrt{1- \left(\frac{q}{2k_F}\right)^2} \left( \frac{1}{e^{\beta_i \hbar c_s q}-1}\right)
\eeq

Which on variable substitution $y = q/2k_F$ and using the definition $k_B T_{BG} = 2 k_F \hbar c_s$\cite{kristen} leads to:

\beq
I(\beta_i) = (2k_F)^4 \int_0^{1} dy \sqrt{1- y^2} \left( \frac{y^3}{e^{y \frac{T_{BG}}{T_i}}-1}\right)
\eeq

Here $\beta_i =\beta$ for phonons and $\beta_i = \beta_e$ for electrons. $T_{BG}$ is called the Bloch-Grueneisen temperature. It corresponds to the maximum phonon wave vector of $2k_F$ (This happens when an electron with wave vector $k_F$ fully scatters exactly backwards generating a phonon with maximum wave vector of $q=2k_F$)\cite{kristen}. $T_{BG}$ is the equivalent of $T_D$ (the Debye temperature in the Debye model of simple metals), and $q=2k_F$ is the equivalent of $q_D$ (the Debye wave vector). The above integral can be simplified in the following two limiting cases and once it is used in equations (18) and (19) we get the following final expressions for the average energy transfer from hot electrons to acoustic phonons in graphene:

\begin{enumerate}

\item The low temperature limit $T_e, T<<T_{BG}$:

\beq
\bar{U} = \frac{3}{\pi^2}\tilde{D} \left(\frac{c_S}{v_F}\right)^2 \left[\frac{T_e^4 -T^4}{T_{BG}^4}\right].
\eeq

\item The high temperature limit $T_e, T>>T_{BG}$:

\beq
\bar{U} = \frac{1}{4\pi}\tilde{D} \left(\frac{c_S}{v_F}\right)^2 \left[\frac{T_e -T}{T_{BG}}\right].
\eeq

Here $\tilde{D} = \frac{D^2 k_F^5}{\rho_m c_s}$. 
\end{enumerate}

These results agree with the reported results in the respective limiting cases\cite{kristen,buts,mohd,caru}.

 \section{Kaganov-Lifshitz-Tanatarov theory for tilted Dirac cone materials}
 
 In this section we work out KLT theory for the tilted Dirac-cone materials. Before proceeding to the calculations, we give a brief introduction to the tilted Dirac matreials and their classifications. These can be further classified into four types based on the tilt nature of the Dirac cones. While considering the case of positive chemical potential, if the Dirac cones are upright, then the Fermi surface is a circle and is called the isotropic or non-tilted Dirac cones (as already considered in the previous section, the case of graphene). If the Dirac cones are slightly tilted and the Fermi surface is still compact but an ellipse, it is called the type-I Dirac cones. Beyond a critical tilt angle, the Fermi surface becomes open and loses its compactness. While the tilt of the Dirac cones is just above the critical tilt angle then the Fermi surface is a parabola and they are identified as type-III Dirac cones. If the tilt angle is much more than the critical tilt angle, the Fermi surface forms a hyperbola and the Dirac cones are identified as type-II Dirac cones\cite{bg0}.  

Graphene is the classic example for the case of isotropic Dirac cones with no-tilt. On the other hand, Hydrogenated Graphene exhibits a type-I tilted Dirac cone\cite{bg1}. Materials like PdTe2 have type-II tilted Dirac cones\cite{bg2} whereas the existence of type-III dirac cones have been reported when Black phosphorus is floquet driven\cite{bg3}.  However, in the current investigation we restrict ourselves to type-I tilted Dirac cone materials. We consider the following toy model for type-I case: 

  \begin{figure}[h!]
    \centering
    \includegraphics[width=0.6\columnwidth]{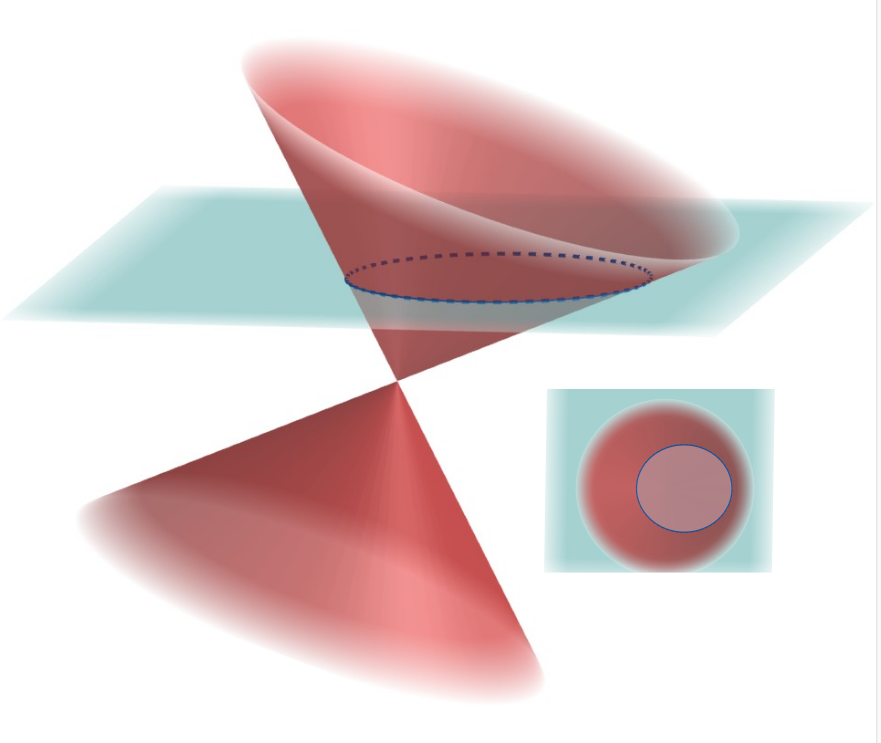}
    \caption{Tilted Dirac cone (Type-I material).}
  \end{figure}
 
\begin{enumerate}

\item For relevant energy scales, electronic dispersion in the tilted Dirac cone is linear in wave vector but: 
\beq
\ep_k = \hbar v_F |\mathk|  + \hbar v_t  |\mathk| \cos(\phi).
\eeq
We assume that Fermi energy is tuned and lies in the conduction band. $v_t$ is the tilt velocity and $\phi$ is the angle between the wave vector $\mathk$ and the direction of the tilt velocity. If the tilt velocity is set to zero, this dispersion goes back to that of graphene. 

\item We assume that the scattering rate due to electron -(acoustic)phonon interaction is given by the same expression\cite{kristen}:
\beq
W_q = \frac{\pi D^2 q}{\rho_m c_s } \left( 1- \left(\frac{q}{2k}\right)^2\right).
\eeq
Here $D$ is the deformation potential constant, $c_s$ is the sound speed, and $\rho_m$ is the mass density $(kg/m^2)$ of the 2D material. 
\end{enumerate}
 
As before, the rate of phonon generation is given by equation (10). Up to equation (16) the calculation proceeds in the similar way. However, in this section we will not preform integration over $k$ as we want to analyze the $k-$space structure of the relaxation dynamics. Equations (18) and (19) now takes the form:

\begin{widetext}
\beq
\bar{U} = C_1 \int d^2k \delta(\ep_k-\mu) \int dq q^4    \left(1- \left(\frac{q}{2k}\right)^2\right) [n_q(\beta_e) - n_q(\beta)] \int_0^{2\pi} d\theta \delta(\ep_{k+q}-\ep_k-\hbar\om_q).
\eeq
\end{widetext}
With  \[~C_1 = \frac{D^2\hbar^2 c_s}{(2\pi)^3\rho_m}.\]
 
Here $\theta$ is an angle between $q$ and $k$ (figure 3), and $\phi$ is the angle between wave vector $\mathk$ and the direction of the tilt velocity (the tilt angle).

  \begin{figure}[h!]
    \centering
    \includegraphics[width=0.6\columnwidth]{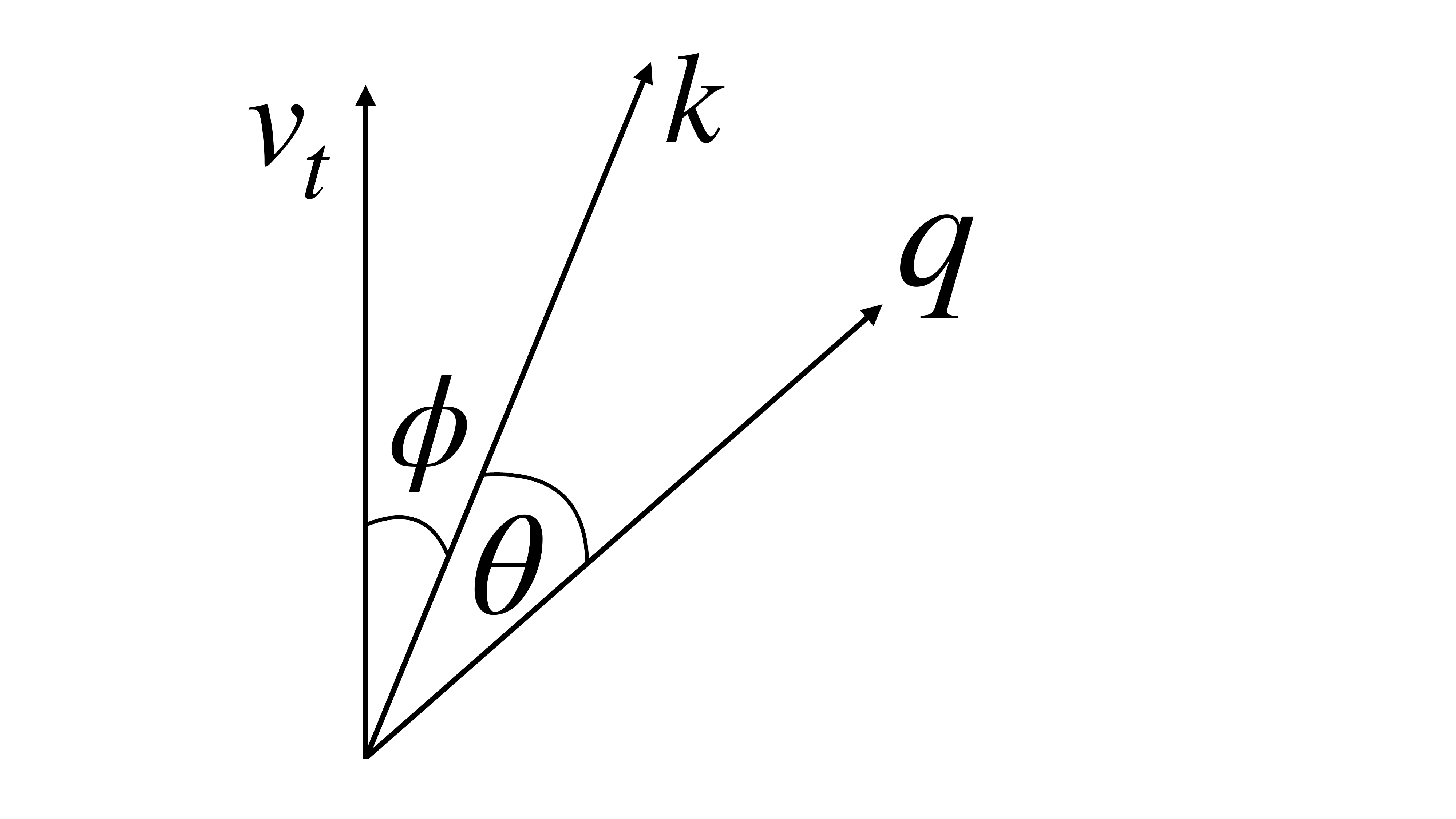}
    \caption{Definition of the angles.}
  \end{figure}
 
In equation (26) we first perform integration over angle $\theta$, and define:

\beq
I(k,q,\phi) = \int_0^{2\pi} d\theta \delta(\ep_{k+q}-\ep_k-\hbar\om_q).
\eeq

Using the definition of the dispersion (equation (25)) the above equation can be written as

\begin{eqnarray}
&&I(k,q,\phi) = \frac{1}{\hbar k v_F (\vtb \cos(\phi) +1)}\nonumber\\
&\times&\int_0^{2\pi} d\theta\delta(\sqrt{1+(q/k)^2  + 2(q/k)\cos(\theta)}-1-\eta(\phi))\nonumber\\
\end{eqnarray}

Where $\eta(\phi)$ is given by

\beq
\eta(\phi) = \frac{\hbar\om_q}{\hbar k v_F (\vtb\cos(\phi)+1)} <<1.
\eeq

As before, we have assumed that phonon energy scale is much smaller than the electronic energy scale. To simplify the delta function we use the standard expression $\delta(f(x)) = \sum_i \frac{\delta(x-x_i)}{|f'(x_i)|}$, where $x_i$ are the roots of $f(x)$. To simplify equation (28) under the said approximation (equation (29)) we need to find the roots of the equation:

\beq
f(\theta) = \sqrt{1+(q/k)^2  + 2(q/k)\cos(\theta)}-1-\eta(\phi)= 0
\eeq

The root in the range $0$ to $2\pi$ is given by $\theta_0 = \cos^{-1}(-\frac{q}{2k})$. This is a valid root as phonon wave vectors are much smaller in magnitude than the electron wave vectors. Thus the inverse cosine function will have a finite and real value. Using this, the integral $I(k,q,\phi)$ simplifies to

\beq
I(k,q,\phi) = \frac{1}{\hbar q v_F (\vtb\cos(\phi)+1)}\frac{1}{\sqrt{1-(q/2k)^2}}
\eeq

Using the expression for $I(k,q,\phi)$, equation (26) now reads

\begin{widetext}
\beq
\bar{U} =\frac{1}{\hbar v_F} C_1 \int d^2k \delta(\hbar v_t k \cos(\phi) +\hbar v_F k-\mu) \int dq q^3 \sqrt{ \left(1- \left(\frac{q}{2k}\right)^2\right)} [n_q(\beta_e) - n_q(\beta)] \frac{1}{\vtb \cos(\phi)+1}.
\eeq
\end{widetext}

Next, the integration over $k$ can be performed $\int d^2 k = \int dk k \int d\phi$. We keep the integration over $\phi$ as we want to analyze the angle dependent relaxation in the Brillouin zone. The integration over the magnitude of $k$ can be immediately performed using the above mentioned property of the delta function: 

\begin{widetext}
\beq
\bar{U} = C \int_0^{2\pi} d\phi \left(\frac{1}{\vtb\cos(\phi)+1}\right)^3 \int_0^{q_{BG}(\phi)} dq q^3 \sqrt{ \left(1- \left(\frac{q}{2k_F(\phi)}\right)^2\right)} [n_q(\beta_e) - n_q(\beta)].
\eeq
\end{widetext}

Here $k_F(\phi)$ is the angle dependent Fermi wave vector due to the presence of the tilt, and is given by

\beq
k_F(\phi) =\frac{\mu}{\hbar v_F (\vtb\cos(\phi)+1)}.
\eeq
And $q_{BG}(\phi) = 2 k_F(\phi)$ is the Bloch-Grueneisen wave vector\cite{kristen}. Here $C = \frac{D^2\mu c_s}{(2\pi)^3\hbar v_F^3 \rho_m}$. 

From equation (33) tilt angle dependent $\bar{U}$ can be defined:

\begin{widetext}
\beq
\bar{U}(\phi) = C \left(\frac{1}{\vtb\cos(\phi)+1}\right)^3 \int_0^{q_{BG}(\phi)} dq q^3 \sqrt{ \left(1- \left(\frac{q}{2k_F(\phi)}\right)^2\right)} [n_q(\beta_e) - n_q(\beta)].
\eeq
\end{widetext}

Such that

\beq
\bar{U} =  \int_0^{2\pi} d\phi \bar{U}(\phi).
\eeq
 
 Expression (35) is one of our main result. We numerically plot $\bar{U}(\phi)$ in figure (4).  We notice the central observation of this work. The energy relaxation is anisotropic in the Brillouin zone. It is maximum for the directions of wave vector $\mathk$ at an angle of $\pi$  with the direction of the tilt velocity, that is, when $\mathk$ is opposite to the director of the tilt velocity (which, in figure (4) is set along the positive x-axis). It is minimum when the wave vector $\mathk$ is along the direction of tilt velocity. 

Next, we integrate over $q$ in various limiting cases. Before that, let us define an effective Bloch-Grueneisen temperature:
\beq
k_B T_{BG} = \hbar c_s \bar{q}_{BG}, ~~~\bar{q}_{BG} = \frac{1}{2\pi}\int_0^{2\pi} d\phi q_{BG}(\phi)
\eeq
$\bar{q}_{BG} = 2k_F(\phi)$ turns out to be

\beq
\bar{q}_{BG} = \frac{\mu}{\pi\hbar v_F}\int_0^{2\pi} d\phi \frac{1}{(\bar{v}_t\cos(\phi)+1)} = \frac{2\mu}{\hbar v_F \sqrt{1-\vtb^2}}
\eeq
And
\beq
k_B T_{BG} = 2\mu (c_s/v_F)\frac{1}{\sqrt{1-\vtb^2}}.
\eeq
By substituting $x=\frac{q}{q_{BG}}$ into equation (35) and after simple algebra, we obtain
\begin{widetext}
\beq
\bar{U}(\phi) = C \left(\frac{2\mu}{\hbar v_F}\right)^4 \left(\frac{1}{\vtb\cos(\phi)+1}\right)^7 \int_0^1 dx x^3 \sqrt{1-x^2} \left[\frac{1}{\exp(\frac{T_{BG}}{T_e}\zeta(\phi) x) -1}  - \frac{1}{\exp(\frac{T_{BG}}{T}\zeta(\phi) x) -1}   \right].
\eeq
\end{widetext}
Here $\zeta(\phi) = \frac{q_{BG}(\phi)}{\bar{q}_{BG}}$, and it is easy to check (using equation (34)) that
\beq
\sqrt{\frac{1-\vtb}{1+\vtb}}<\zeta(\phi)<\sqrt{\frac{1+\vtb}{1-\vtb}}
\eeq

 \begin{figure}[h!]
    \centering
    \includegraphics[width=1.0\columnwidth]{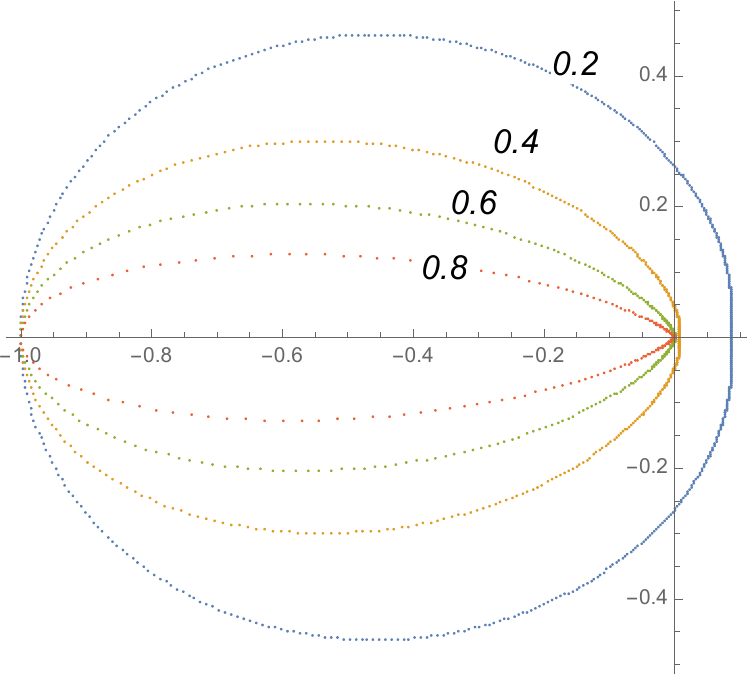}
    \caption{Polar plot of $\bar{U}(\phi)$ for various values of scaled tilt velocity $\vtb=\frac{v_t}{v_F} = 0.2, 0.4, 0.6, 0.8$, as depicted in the figure. We notice that the energy relaxation is anisotropic in the Brillouin zone. It is maximum for the directions of wave vector $\mathk$ at an angle of $\pi$  with the direction of the tilt velocity. The tilt velocity points towards the positive x-direction (in the figure). $\bar{U}(\phi)$ is minimum when the wave vector $\mathk$ is along the direction of tilt velocity. For the numerical plot we took: $\mu = 2~eV$, lattice temperature $T= 100~K$, electron temperature $T_e = 400~K$, $v_F = 10^{6}~m/s$, sound speed $c_s = 10^{3}~m/s$.}
  \end{figure}

Next, let us consider the limiting cases one by one.:

\subsection{CASE A: $T,~T_e<<T_{BG}$} 

In this low temperature limit, the factor in the exponent $\frac{T_{BG}}{T_e}\zeta(\phi) x$ in equation (40) is much much greater than one, given the bound on $\zeta(\phi)$ (assuming $v_t<v_F$). The Bose factors can be approximated as 

\[\frac{1}{\exp(\frac{T_{BG}}{T_e}\zeta(\phi) x) -1} \sim e^{-\frac{T_{BG}}{T_e}\zeta(\phi) x}  \]

Integrations can be easily performed, and we obtain

\begin{widetext}
\beq
\bar{U}(\phi) = 6 C \left(\frac{2\mu}{\hbar v_F}\right)^4 \left(\frac{1}{\vtb\cos(\phi)+1}\right)^3 \frac{1}{(1-\vtb^2)^2} \left[\frac{T_e^4-T^4}{T_{BG}^4}\right]
\eeq
\end{widetext}

We again get back $T^4$ behaviour. However, now the expression for $\bar{U}(\phi)$ has explicit dependence on angle $\phi$ and it is anisotropic in the Brillouin zone (as depicted in figure 4). Next we consider the opposite limit case:

\subsection{CASE B:  $T,~T_e>>T_{BG}$}
In this limit we have $\frac{T_{BG}}{T_e} \zeta(\phi)x<<1$, and a similar expression involving $T$ instead of $T_e$ holds true. The Bose factors can be approximated as

\[\frac{1}{\exp(\frac{T_{BG}}{T_e}\zeta(\phi) x) -1} \sim \frac{T_e}{T_{BG}}\frac{1}{\zeta(\phi)x}. \]

The expression (40) now leads to

\beq
\bar{U}(\phi) = \frac{\pi}{16} C \left(\frac{2\mu}{\hbar v_F}\right)^4 \left(\frac{1}{\vtb\cos(\phi)+1}\right)^6 \frac{1}{\sqrt{1-\vtb^2}} \left[\frac{T_e-T}{T_{BG}}\right]
\eeq

The final expressions for $\bar{U}$ integrated over $\phi$ can be easily obtained from equations (42) and (43), using the equation (36).

Next, we would like to explore the question: What is the experimental relevance of the obtained results (equations (40), (42), and (43))? To this end we consider an example of a type-I tilted Dirac cone material: 8-Pmmn Borophene. This compound is proposed to be a type-I tilted Dirac material, but it is not yet synthesized\cite{boro}. As shown in the figure (5),  8-Pmmn Borophene crystallizes in an orthorhombic structure with 8 Boron atoms per unit cell. The tilt velocity is directed towards the $b$-direction in the real space of the crystal structure, as shown in the figure. In the direction of the tilt velocity, the value of $\bar{U}(\phi)$ is $\bar{U}(0)$ as the angle between wave vector $\mathk$ and the tilt velocity is zero. In the opposite direction it is $\bar{U}(\pi)$. Therefore, the average rate of energy relaxation along the $b$-direction (figure (5)), or along the direction of "directly connected Boron ridge atoms" is given as:

\[U_{ridge} = \bar{U}(0) + \bar{U}(\pi)\]

 \begin{figure}[h!]
    \centering
    \includegraphics[width=0.8\columnwidth]{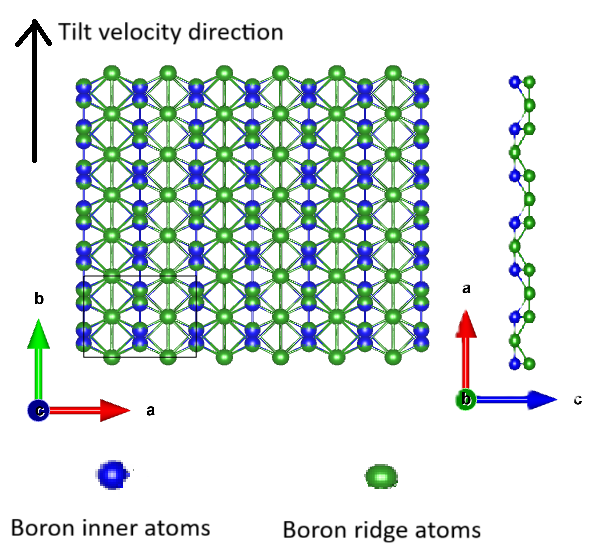}
    \caption{8-Pmmn Borophene: a type-I tilted Dirac cone material.}
  \end{figure}

In the perpendicular direction (that is, along the $a-$direction) the average rate of energy relaxation is given as:

\[U_{\perp} = \bar{U}\left(\frac{\pi}{2}\right) + \bar{U}\left(\frac{3\pi}{2}\right)\]

From equation (40) the numerical value of the ratio
\beq
Ratio = \frac{U_{ridge}}{U_{\perp}}
\eeq
turns out to be about $32$ for $\vtb =0.5$ and other parameters as mentioned in figure (4).  Thus, we have this important observation: along the direction of ridge atoms (b-direction) more phonons will be generated, and this direction will experience more heating. The perpendicular direction (a-direction) lesser phonons will be generated. The most important point to be kept in mind is that this heating will persist only at pico-second time scales (the time scales of phonon relaxation). Ideally, one should consider direction specific phonon branches and perform this investigation by going beyond the TTM (uniform phonon temperature). This is kept as an open problem for future investigation.

In general, the ratio (equation (44)) is tilt velocity dependent:
\beq
Ratio(\vtb) = \frac{U_{ridge}(\vtb)}{U_{\perp}(\vtb)},
\eeq
and using equation (40) it is computed for various values of the tilt velocity and plotted in figure (6).

 \begin{figure}[h!]
    \centering
    \includegraphics[width=0.8\columnwidth]{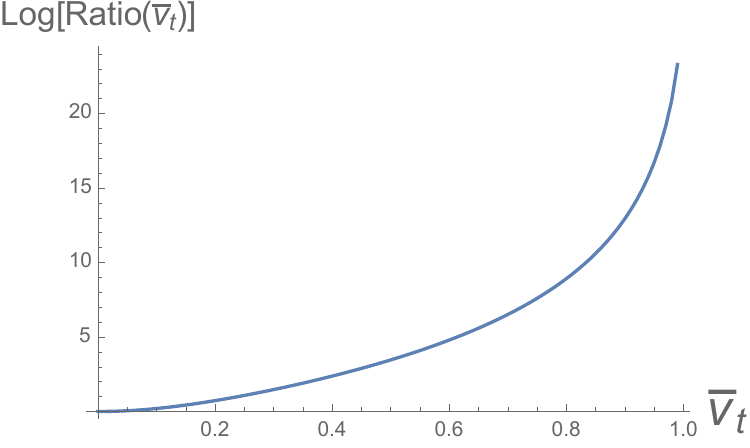}
    \caption{Log of the $Ratio(\vtb)$ as a function of $\vtb$ for the same parameters as given in the caption of figure (4)}
  \end{figure}

It is to be noted that this ratio exponentially increases as the scaled tilt velocity is increased.  The direction of the tilt velocity can be externally controlled (say for example, via application of external pressure\cite{bg0}). Therefore, the discovered physical effect will have important experimental consequences such as the magnitude of the transient Seebeck effect (or the magnitudes of the transient temperature gradients) can be externally controlled with pressure.

\section{Conclusion}

We observe that in the type-I tilted Dirac cone materials the non-equilibrium (hot) electron relaxation with phonons is anisotropic. It means that there is a preferential heating of the lattice degrees of freedom in the specific directions of the Brillouin zone, in particular, in the direction opposite to the tilt velocity.  This uncovered physical effect will have non-trivial experimental consequences: (1) With pump-probe spectroscopy on a given tilted Dirac cone material an anisotropic relaxation would lead to anisotropic heating which can further lead to a transient Seebeck effect as transient thermal gradients would exist, and (2) the direction of anisotropic heating can be controlled externally by controlling the direction of the tilt velocity which can be tuned by the application of an external pressure. Therefore, we foresee novel consequences of this physical effect.

\vspace{1cm}

\section{Acknowledgements}
S Mandal acknowledges the funding from APEX project(Institute of Physics, Bhubaneswar), theoretical, travel head.

\section{Appendix}

With linear dispersion $\ep_k = \hbar v_F k$, the integral over $k$ in equation (13) can be written as

\begin{widetext}
\beq
\int d^2k (f(\ep_k) -f(\ep_k) - \hbar \om_q \frac{\pr f}{\pr \ep_k} -....)\delta(\hbar v_F \sqrt{k^2+q^2 + 2 k q\cos(\theta)} -\hbar v_F k -\hbar \om_q).
\eeq

As the Fermi energy is the largest scale one can write $-\frac{\pr f}{\pr \ep_k} \simeq \delta(\ep_k-\mu)$. Thus the above expression takes the form

\beq
\frac{1}{(\hbar v_F)^2} \int_0^\infty dk \int_0^{2\pi} d\theta \delta(k-k_F)\delta(\sqrt{1+q^2/k_F^2 + 2 (q/k_F)\cos(\theta)} - 1 - (\hbar \om_q)/(\hbar v_F k_F)).
\eeq
\end{widetext}

Integral over $k$ can be immediately done. To simplify the delta function we use the standard expression $\delta(f(x)) = \sum_i \frac{\delta(x-x_i)}{|f'(x_i)|}$, where $x_i$ are the roots of $f(x)$.  The valid root turns out to be $\theta_i = \cos^{-1}(c_s/v_F - q/(2k_F))$, and integrating over $\theta$ leads to equation (17).



\end{document}